\documentclass[aip,jap,showpacs,amsmath,amssymb,reprint]{revtex4-1}
\usepackage[utf8]{inputenc}
\usepackage{graphicx}
\usepackage{esint}
\usepackage{epstopdf}
\DeclareGraphicsExtensions{.eps,.png}

\begin{document}
\newcommand{\meff}{m_{\mathrm{eff}}}
\newcommand{\jav}{j_{\mathrm{av}}}
\newcommand{\icmin}{i_{c,{\mathrm{min}}}}

%\author{Lior Ella\inst{1} \and D. Yuvaraj \and Oren Suchoi \and Oleg Shtempluk \and Eyal Buks\inst{1}}

%\institute{\inst{1} Department of Electrical Engineering, Technion - Haifa 32000 Israel}

\title{Tunable strong nonlinearity of a micromechanical beam embedded in
a dc-SQUID}

\author{Lior Ella}
\altaffiliation{Present address: Weizmann Institute, Rehovot 7610001 Israel}
\email{lior.ella@weizmann.ac.il}

\author{D. Yuvaraj}
\altaffiliation{Present address: London Centre for Nanotechnology,
University College London,
London WC1H 0AH.}

\author{Oren Suchoi}

\author{Oleg Shtempluk}

\author{Eyal Buks}

\affiliation{Faculty of Electrical Engineering, Technion, Haifa 32000 Israel}
\begin{abstract}
We present a study of the controllable nonlinear dynamics of a micromechanical beam coupled to a dc-SQUID (superconducting quantum interference device). The coupling between these systems places the modes of the beam in a highly nonlinear potential, whose shape can be altered by varying the bias current and applied flux of the SQUID. We detect the position of the beam by placing it in an optical cavity, which frees the SQUID to be used solely for actuation. This enables us to probe the previously unexplored full parameter space of this device. We measure the frequency response of the beam and find that it displays a Duffing oscillator behavior which is periodic in the applied magnetic flux. To account for this, we develop a model based on the standard theory for SQUID dynamics. In addition, with the aim of understanding if the device can reach nonlinearity at the single phonon level, we use this model to show that the responsivity of the current circulating in the SQUID to the position of the beam can become divergent, with its magnitude limited only by noise. This suggests a direction for the generation of macroscopically distinguishable superposition states of the beam.
\end{abstract}

\pacs{85.25.Dq, 85.85.+j, 05.45.-a}

\maketitle

\section{Introduction}

Micro and Nano-Electromechanical systems (NEMS and MEMS) have been a subject of intense research in the past decade~\cite{poot_mechanical_2012,ekinci_ultimate_2004,lifshitz_nonlinear_2008,karabalin_nonlinear_2009,kozinsky_basins_2007,lahaye_nanomechanical_2009,teufel_sideband_2011,groblacher_radiation-pressure_2008}, due to their potential for both probing fundamental physical questions, such as the limits of validity of quantum mechanics~\cite{teufel_circuit_2011,lahaye_nanomechanical_2009,agarwal_witnessing_2012}, and for functioning as highly sensitive, quantum-limited detectors~\cite{buks_displacement_2007,ekinci_nanoelectromechanical_2005,etaki_dc_2011,etaki_motion_2008,poot_mechanical_2012}. One of the appealing aspects of these devices is their tendency to display nonlinear behavior. This, in addition to providing an experimentally accessible testbed for studies of nonlinear dynamical systems~\cite{schwab_spring_2002,kozinsky_basins_2007,karabalin_nonlinear_2009,dykman_fluctuating_2012,lifshitz_nonlinear_2008,defoort_scaling_2014,villanueva_surpassing_2013,kenig_optimal_2012}, is a resource for the generation of nonclassical states of mechanical elements~\cite{Yurke_13,yurke_generating_1986,katz_signatures_2007,katz_classical_2008,ludwig_optomechanical_2008,qian_quantum_2011,Xue_064305}.

A particular type of nonlinearity, that of a resonator with an amplitude-dependent spring constant (Duffing resonator), can be gainfully harnessed for this end: It has been shown that both the multi-phonon transitions it exhibits, as well as its inherent bistability, enable the generation of a superposition of macroscopically distinct coherent states~\cite{Yurke_13,katz_signatures_2007,katz_classical_2008}. It is therefore highly advantageous to be able to generate Duffing nonlinearity in NEMs and MEMs which is both strong and can be controlled, tuned, and detected by the experimenter.

In this work we demonstrate the possibility to achieve such a controllable nonlinearity in a mechanical beam embedded in a dc-SQUID and placed in an external magnetic field. The magnetomotive interaction of the SQUID with the beam places the latter in a highly nonlinear potential, which in particular gives rise to a Duffing nonlinearity. The shape of the potential, and with it the resonance frequency and Duffing coefficient of the beam modes, can be altered by varying the control parameters (bias current and applied bias flux) of the SQUID. 

In previous work on a similar system~\cite{etaki_motion_2008,poot_tunable_2010,etaki_dc_2011,schneider_coupling_2012,etaki_self-sustained_2013}, the SQUID was used both to read out the position of the beam in addition to influencing its dynamics. As a result, the SQUID could only be biased at an operating point in which the voltage is sufficiently dependent on the flux to allow displacement detection. While this scheme provided a highly sensitive displacement measurement, it also placed a restriction on the range of control parameters that could be explored. In contrast, in our work displacement detection is independent of the SQUID, which enables us to explore the full space of control parameters of the device.

In our device, displacement detection is obtained by forming an optical cavity between the beam and the tip of an optical fiber placed directly above it~\cite{zaitsev_forced_2011} (see Fig.~\ref{fig:A-schematic-of}). The cavity is driven by a laser, and power reflected off of it is dependent on the displacement of the beam. To actuate the beam, we coat the tip of the fiber with Niobium and apply a biased AC voltage, which drives the beam capacitively (see Fig.~\ref{fig:A-schematic-of}). Using this scheme, we measure the frequency response of the fundamental beam mode near resonance, from which we extract the dependence of its resonance frequency and Duffing coefficient~\cite{kozinsky_basins_2007,lifshitz_nonlinear_2008} on the control parameters of the SQUID.

Interestingly, we find that the resonance frequency and Duffing coefficient display pronounced periodic oscillations as the bias flux of the SQUID is varied (see Figs.~\ref{fig:Comparison-of-frequecy-low} and~\ref{fig:Comparison-of-frequecy-hi}), which can be directly attributed to the flux-periodic response of the SQUID. These oscillations change their shape as the bias current is varied, and their magnitude is largest near the transition from the zero voltage state (S-state) of the SQUID to its resistive state (R-state). A model, based on the standard theory of SQUID dynamics (RCSJ), is developed which accounts for the results. While most of the qualitative as well quantitative details of the measurements are reproduced by this model, several discrepancies exist, as shown in Fig.~\ref{fig:Frequency-response-s-state}. 

A specific and previously unattainable bias point of the SQUID, for which the nonlinearity is expected to be particularly strong, is at the transition to the resistive state when the bias flux is set at half-integer values in units of the magnetic flux quantum.  We argue that as the bias current and applied bias flux of the SQUID approach this point, the induced resonance frequency shift and Duffing coefficient of the beam diverge, and that this divergence is physically limited by noise in the SQUID. Since this transition is in fact an infinite period bifurcation~\cite{strogatz_nonlinear_2001}, in what follows we shall refer to this point as the bifurcation cusp point.

\section{The experiment}

\subsection{\label{sub:Overview-of-the-system}Overview of the system}

\begin{figure}
\begin{centering}
\includegraphics[width=1\columnwidth]{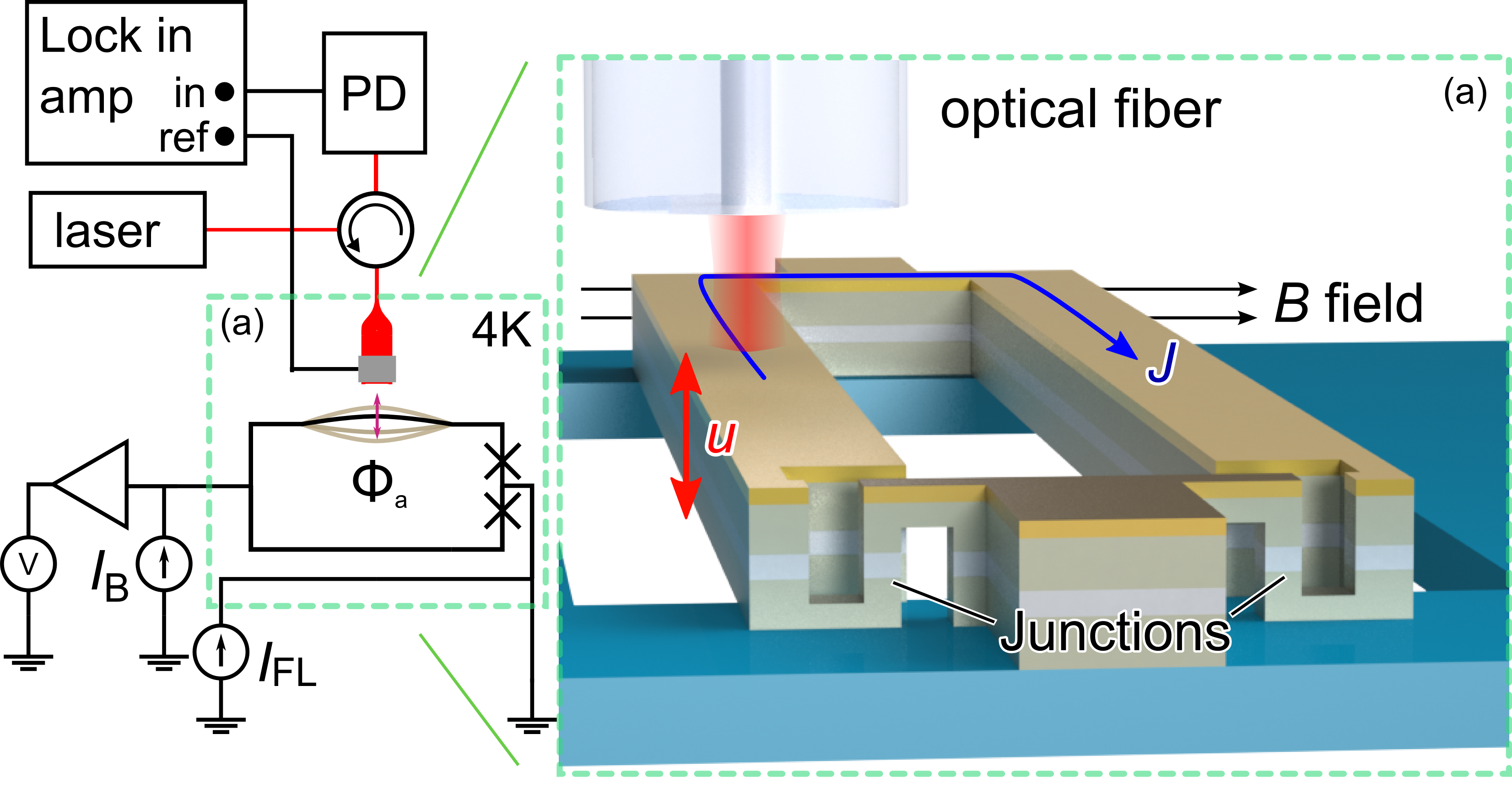}
\par\end{centering}

\caption{(color online) A schematic description of the experimental system. The SQUID is biased
with a current $I_{B}$, and the voltage across it is amplified and
measured. The displacement of the mechanical modes, which are placed in a transverse magnetic field, is detected with
an optical cavity. This cavity is formed by the beam on one side and the tip of the optical
fiber (red) on the other. The fiber is coated with Niobium electrode,
which is set, using a bias tee (not shown) at a finite dc voltage and connected to the reference output of an RF lock
in amplifier. The power reflected from the cavity is converted
to voltage with an RF photodetector, and then measured with the lock in amplifier.\label{fig:A-schematic-of} (a): A 3D blow-up of the SQUID, with the junctions shown in front.
The fiber is located above one of the beams. The displacement of the measured beam mode is denoted by $u$, and the circulating current in the SQUID by $J$. The SQUID is fabricated in a Nb/AlOx/Nb configuration, and is top-coated with gold.}
\end{figure}

%rewrite notes:
%1. it can't start from this. Needs to add some kind of introduction that describes the device (sentence or two). The details should be condensed.
The device was created by patterning a dc-SQUID in a trilayer configuration on a SiN coated Si substrate~\cite{yuvaraj_fabrication_2011}. A part of the SQUID loop was freed and suspended in vacuum, and functioned as a mechanical beam. The displacement of this beam was detected by placing an optical fiber above, which forms a cavity between the top of the SQUID and the fiber tip (see Fig.~\ref{fig:A-schematic-of}). While two beams were freed, our experiment focused on the dynamics of the fundamental mode of only one of them.
The Josephson junctions (JJs), which were overdamped and non-hysteretic, were found to have an average critical
current $I_{0}=317.5{\rm \mu A}$ at zero magnetic field.
Further details regarding the SQUID, and definitions of
SQUID parameters used subsequently for modeling the dynamics of the device,
can be found in the appendices.
The mechanical elements functioned as doubly-clamped beams of length
$\ell=100{\rm \mu m}$. We measured the frequency response of the fundamental
mode of one of the beams, which had an angular frequency $\omega_{0}=2\pi\times 311.75{\rm KHz}$
and quality factor $Q_{m}\simeq 6200$. The system was placed in an
external magnetic field of $60{\rm mT}$ formed by a split-coil magnet. The field was aligned with the plane of the sample, although a small component perpendicular to the plane of the SQUID existed and contributed to the flux threading the loop.

\subsection{\label{sub:Experiment,-results-and}Experiment and results}

\begin{figure*}
\includegraphics[width=1\textwidth]{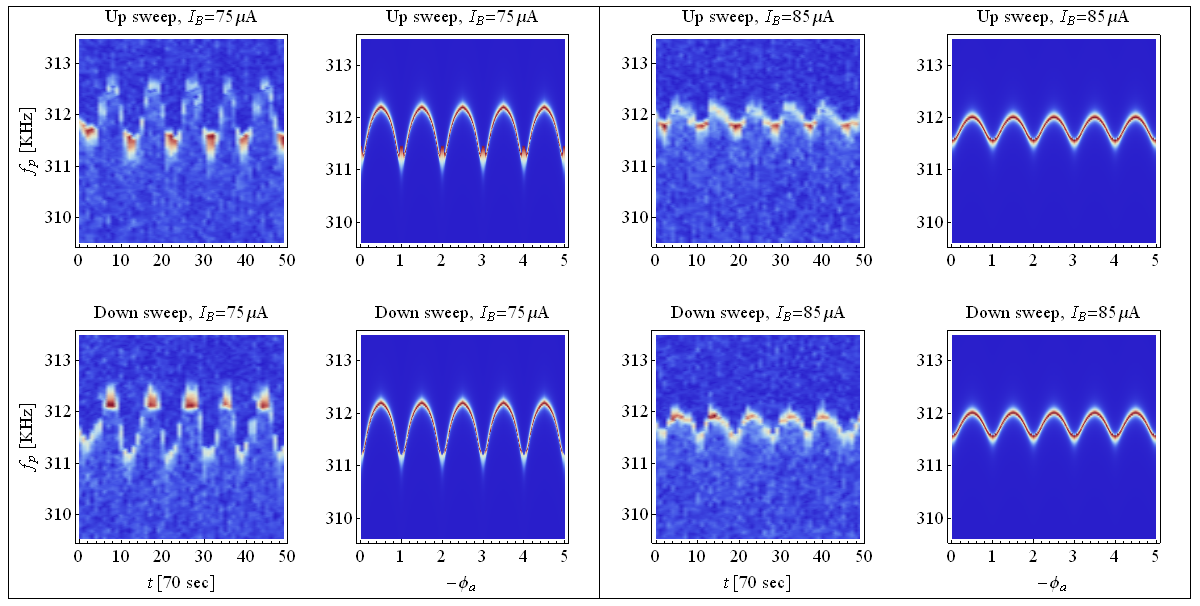}

\caption{\label{fig:Comparison-of-frequecy-low}Comparison of frequency response
measurements to the theoretical model.
The left panel in each frame shows the experimental
measurement and the right panel shows the theoretical prediction,
obtained with the model outlined in section~\ref{sub:Experiment,-results-and}. The up (down) sweeps correspond to the direction in which the frequency of the signal actuating the beam was altered.
The abscissa is $-\phi_{a}$ in the theoretical panel since in the experiment
the flux decreased with time. Blue (red) colors denote a weaker (stronger)
response.
The plots here show the frequency response when the system is fully
in the R-state, for bias currents $I_{B}>2I_{0}$. The sinusoidal modulation in the resonance frequency of the beam and the hysteretic tilted Lorentzian response, which corresponds to a Duffing nonlinearity, are induced by the SQUID. 
}
\end{figure*}

The influence of the SQUID on the beams was measured by obtaining
the frequency response of the beams to a sinusoidal capacitive force
near the resonant frequency of the fundamental mode. In the absence
of the split coil magnetic field, the response of the mode was independent
of the SQUID bias current $I_{B}$ and the applied flux
$\Phi_{a}$. When the field was turned on and the bias currently was increased, the frequency response developed a pattern which had unique features for different values of $I_B$, which were periodic in the applied flux. (see 
Figs.~\ref{fig:Comparison-of-frequecy-low},~\ref{fig:Comparison-of-frequecy-hi}
and~\ref{fig:Frequency-response-s-state}).
The features were most pronounced near the transition from the S-state to the R-state of the SQUID, and subsequently began to decay as $I_B$ was further increased to the regime in which the SQUID displayed ohmic behavior.
Note that the $\Phi_a$ was sweeped by allowing the magnetic field in the split coil magnet to freely decay and making use of the imperfect alignment of the field with the plane of the sample~\cite{schneider_coupling_2012}.

At $I_B=0$ the response of the beam mode to actuation could be fitted to a Lorentzian, indicative of a a harmonic response. As $I_B$ was increased, however, the response started to exhibit, in addition to a resonance frequency shift, a ``tilted'' Lorentzian characteristic of a Duffing oscillator. To verify this, the response was swept both in the up and down directions, and hysteretic response, indicative of a Duffing bistability, was clearly observed. Sharp transitions in the up and down sweeps correspond to an amplitude-dependent spring hardening and softening, respectively. For some values of control parameters, the hysteretic behavior was particularly pronounced, indicating a strong nonlinearity of the beam mode.

\subsection{Discussion and theory}
To understand the observed frequency response, we first outline the dynamics of a SQUID coupled to a vibrating beam~\cite{buks_decoherence_2006,blencowe_quantum_2007,poot_tunable_2010,etaki_self-sustained_2013}.
We denote the current in the arms of the SQUID by $I_{n}=I_{0,n}\sin\gamma_{n}$,
where $n=1,2$, $I_{0,n}$ is the critical current in the $n$'th
junction, and $\gamma_{i}$ is the gauge invariant phase across the
junctions.
Furthermore, denoting the component of the applied magnetic field in the plane
of the SQUID as $B$, a Lorentz force $F_{L}=\lambda J\ell B$ acts
on the beams, where $\lambda$ is a correction factor accounting for the mode shape (see appendix and~\cite{nation_quantum_2008,blencowe_quantum_2007})
and $J=(I_{1}-I_{2})/2$ is the circulating current in the SQUID. Concurrently,
the total flux $\Phi$ threading the SQUID is dependent on the displacement
of the beams. To first order, we have $\Phi=\Phi_{a}+\lambda B\ell x+LJ$,
where $x$ is the displacement of the driven mechanical mode from
its equilibrium position, $\Phi_{a}$ is the applied flux threading
the SQUID loop at $x=0$ and $L$ is the self inductance of the loop.
Since $x\ll\ell$ and $\Phi_{a}\gg LJ$,
we can make the approximation $\dot{\Phi}\simeq\lambda B\ell\dot{x}+L_{0}\dot{J}$,
where $L_{0}$ is the loop inductance when the beams are in their
equilibrium positions. 

%this should say something like: We know that the beam is placed in some kind of effective potenital. SO we eliminate the fast squid coords. to obtain the force on the beam mode, which has a Lorentz force component but depends only on the average circulating current.
Since the Lorentz force acting on the mode depends on its displacement, it is placed in an potential whose shape depends on the control parameters of the SQUID. By measuring the mechanical resonance frequency shift and Duffing nonlinearity, the observed frequency response allows us to extract the quadratic and quartic terms of this potential around the equilibrium point. To calculate the Lorentz force acting on the beam, we find the circulating current in the SQUID for the given control parameters, and assume that the mechanical displacement is a small perturbation of the applied flux. Since the oscillation frequency of the SQUID $\omega_c=2\pi R I_0/\Phi_0 \gg \omega_0$, we only need to consider the dc component of the circulating current.

We assume that the equation of motion for the amplitude of the driven mode,
in normalized units, is given by 
\begin{equation}
\ddot{u}+\kappa_{m}\dot{u}+\omega_{0}^{2}u=g^{2}\jav(\phi_{a}+u,i_{B})+h_{d}\cos\omega_{p}t,\label{eq:eom_beam_mode}
\end{equation}
%explain what j_av is.
where $u=x/x_{B}$, overdot denotes time derivative, $\kappa_{m}=\omega_{0}/Q_{m}$,
$g=\lambda\ell B\sqrt{I_{0}/m_{{\rm eff}}\Phi_{0}}$, $i_{B}=I_{B}/I_{0}$, $I_B=I_1+I_2$,
$\phi_{a}=\Phi_{a}/\Phi_{0}$ is the normalized applied flux, $h_{d}$
is the normalized driving strength and $\omega_{p}$ is the driving
signal angular frequency. Here $x_{B}=\Phi_{0}/\lambda\ell B$ is
the displacement required to change the applied flux by $\Phi_{0}=h/2e$,
$m_{{\rm eff}}$ is the effective mass of the mode, 
% some words about what is j_av are important.
and $\jav=J_{{\rm av}}/I_{0}$
is the averaged and normalized circulating current. In the S-state
$\jav$ is determined by the location of
the stable equilibrium points (wells) of the SQUID potential, and
in the R-state it is given by $\jav=\Theta^{-1}\int_{0}^{\Theta}j(t){\rm d}t$,
where $\Theta$ is a single period of $j(t)=J(t)/I_{0}$. The coordinate
$u$ can be treated adiabatically when solving for the dynamics of
the SQUID since the latter is overdamped and $g^{2}/\omega_{c}\omega_{0}\ll Q_{m}^{-1}$~\cite{schwab_spring_2002,poot_tunable_2010}, where
$\omega_{c}$ is the JJ oscillation frequency at the R-state. Since the SQUID dynamics are highly nonlinear and in the R-state no stable equilibrium points exist, the
general analytical calculation of $\jav$ in both states is difficult, and so we
obtain it numerically (see appendix~\ref{sub:Modeling-the-SQUID-beam-1}). We then find the mode frequency shift and Duffing coefficient
by assuming that $u\ll 1$ and expanding $\jav$ in powers of $u$.

\begin{figure*}
\includegraphics[width=1\textwidth]{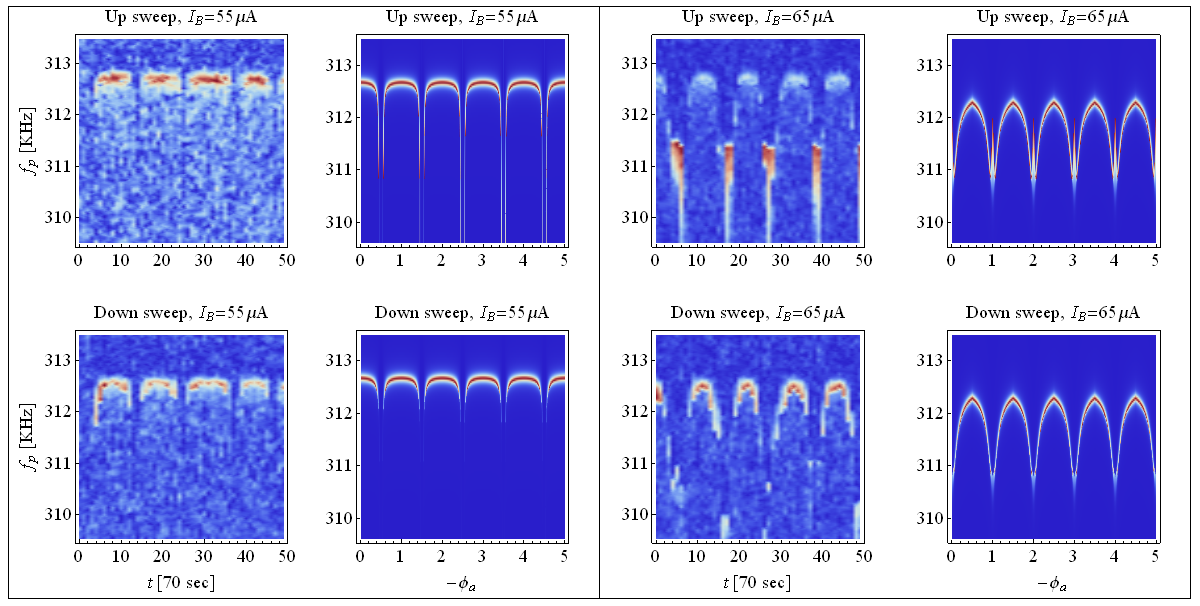}

\caption{\label{fig:Comparison-of-frequecy-hi}(Continued from Fig.~\ref{fig:Comparison-of-frequecy-low}.)
Frequency response lower values of SQUID bias current. The average critical current of each JJ with applied magnetic field
is $I_{0}=30\pm2{\rm \mu A}$, and since $\beta_{L}=4$ in magnetic
field,
we have $I_{c,{\rm min}}=49\pm3{\rm \mu A}$.
Thus, in the plot at
$I_{B}=55{\rm \mu A}$, the SQUID is in the S-state most of the time,
and the sharp dip in the frequency of the beam corresponds to the
bifurcation cusp point.}
\end{figure*}

\begin{figure}
\includegraphics[width=1\columnwidth]{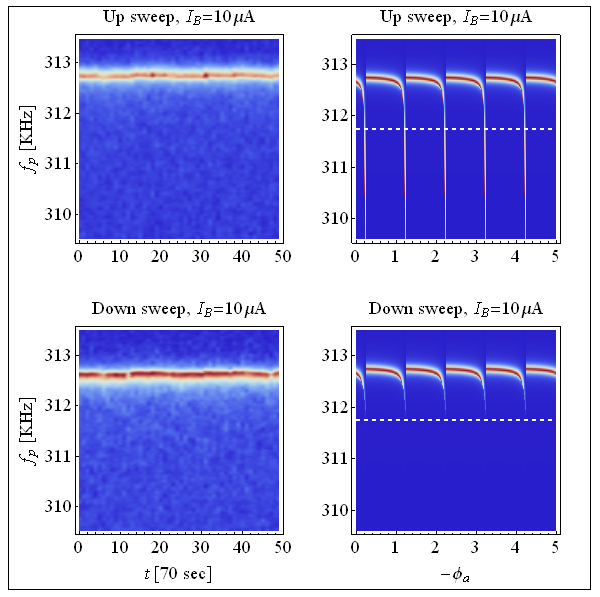}

\caption{\label{fig:Frequency-response-s-state}(color online) Frequency response of the driven
mechanical mode to capacitive actuation, when the bias current $I_{B}$
is smaller than $I_{c,{\rm min}}$. In this case the SQUID is in the
S-state for all values of $\phi_{a}$.
The white dashed line at $f_{p}=311.75{\rm KHz}$ corresponds to the
resonance frequency of the mechanical mode for the case $B=0$. The discrepancy between theoretical prediction and experimental
measurements is discussed in subsection~\ref{sub:Experiment,-results-and}.
Response for higher bias currents is shown in Figs.~\ref{fig:Comparison-of-frequecy-low}
and~\ref{fig:Comparison-of-frequecy-hi}. }
\end{figure}

We can see that above the S-state, in Figs.~\ref{fig:Comparison-of-frequecy-low} and~\ref{fig:Comparison-of-frequecy-hi},
the predicted frequency shift follows the experimental data closely.
However, the Duffing nonlinearity is only in partial agreement
with the data. For example, in Fig.~\ref{fig:Comparison-of-frequecy-hi},
for $I_{B}=75{\rm \mu A}$, the nonlinearity appears to be symmetric,
while the theory suggests that it should be observable only at integer
flux quanta. A larger discrepancy between theory and experiment is
found in Fig.~\ref{fig:Frequency-response-s-state}, which is for
a low bias current, for which the SQUID is in the S-state for all
values of $\phi_{a}$. For $I_{B}<I_{c,{\rm min}}$, the minimal critical
current, the SQUID potential has a multiplicity of stable wells. As
$\phi_{a}$ is varied, these wells disappear and reappear periodically.
The theoretical prediction is that the force on the beams due to circulating
current is approximately linear for most values of $\phi_{a}$, except
near those points in which a well in the SQUID potential disappears.
Thus, the Lorentz force acting on the beams should be linear except
near values of $\phi_{a}$ in which a dip in mechanical frequency
should occur. The measured frequency response, however, does not exhibit
these dips.

Note that an important consequence of the model described by Eq.~(\ref{eq:eom_beam_mode}),
is that $\jav$ is a function of the sum $\phi_{a}+u$. Due to this,
the sign and magnitude of the Duffing coefficient should be proportional
to the second derivative of the frequency shift. This feature is qualitatively
consistent with the experimental data shown in the panels of Figs.~\ref{fig:Comparison-of-frequecy-low}
and~\ref{fig:Comparison-of-frequecy-hi}.

\section{Dynamics near the bifurcation cusp point}

\subsection{\label{sub:Operation-near-the-cusp-pt}Maximal attainable nonlinearity}
% say somehting about what we are going to discuss more clearly.
Since we have seen that very strong nonlinearity is exhibited in this device, it is interesting to consider for which values of the control parameters this effect is most pronounced. To address this question, we consider
a symmetric dc-SQUID with $\beta_{L}\lesssim1$ and normalized capacitance
$\beta_{c}\ll1$, which makes the analysis tractable without changing the results qualitatively.
The normalized bias current $i_{c}(\phi_{a})$ for
which a transition to the R-state occurs is a periodic function of
the normalized applied flux $\phi_{a}$ with period 1, and its minimal
value $\icmin$ occurs at $\phi_{a}=\frac{1}{2}+n$, where $n$ is
an integer. Setting $\delta\phi=\phi_{a}-\frac{1}{2}$, and $\delta i=i_{B}-i_{c,{\rm min}}$,
we focus on the dynamics of the SQUID close to the bifurcation cusp
point $\delta\phi=0$, $\delta i=0$. When the SQUID is biased near
this point, the circulating current $\jav$ becomes extremely sensitive
to the applied flux since for $\delta\phi>0$ ($\delta\phi<0$) it
is energetically more favorable for $\jav$ to be large and negative
(positive), and so the point $\delta\phi=0$ exhibits a singularity
which remains also in a modestly asymmetric SQUID. In the R-state,
the jump in $\jav$ must occur on a span of $\delta\phi$ which is
on the order of $\delta i$. From this we may anticipate that $\partial^{n}\jav/\partial\phi_{a}^{n}\propto(\delta i)^{-n}$.
To verify this, we calculate $\jav$ for $|\delta\phi|\ll1$ and $0<\delta i\ll1$.
Assuming $\beta_{L}\ll1$, we may use adiabatic elimination to set
$j=-\cos\left(\frac{\gamma}{2}\right)+O(\beta_{L})$, where $\gamma=\gamma_{1}+\gamma_{2}$,
and reduce the dynamics near $\phi_{a}=\frac{1}{2}$ to the one-dimensional
equation ${\rm d}\gamma/{\rm d}\tau=-{\rm d}v/{\rm d}\gamma+O(\beta_{L}^{2}),$
where 
\begin{equation}
v(\gamma)=4\pi\delta\phi\cos\left(\frac{\gamma}{2}\right)-\frac{1}{2}\pi\beta_{L}\cos\left(\gamma\right)-i_{B}\gamma\label{eq:gamma_pot}
\end{equation}
and $\tau=\omega_{c}t$. This equation describes overdamped motion
of $\gamma$ in a ``double'' washboard potential. When $0<\delta i\ll1$
and $|\delta\phi|\ll1$, this potential no longer contains any wells.
It does, however, contain nearly flat regions around the points $\gamma_{c}$,
defined by $v''(\gamma_{c})=0$ and $v'''(\gamma_{c})<0$, in which
the dynamics are slow. In fact, during a single period $\Theta$ of
$j$, the time spent away from these points scales as $\sqrt{\delta i}$,
and so it is sufficient to solve for the dynamics around them.

Restricting our attention to $-2\pi\le\gamma\le2\pi$, a single period of $v(\gamma)+i_{B}\gamma$,
we have two such points, which we denote as $\gamma_{c\pm}$. If we
expand the potential around them and keep terms up to quadratic
order, we may solve the resulting equations and find an approximate
analytical expression for $\jav$, which is correct up to an error
of $O(\sqrt{\delta i})$. A plot of $\jav$ obtained using this analytical
expression is given in Fig\@.~\ref{fig:The-average-circulating},
and its explicit form can be found in appendix~\ref{sec:Analytical-expression-for-jav}.
Expanding this expression around $\delta\phi=0$, and making the additional
assumption that $\delta i\ll\beta_{L}$, we find that
\begin{equation}
j_{{\rm av}}(\delta\phi)=\frac{\pi}{2\sqrt{2}}\frac{\delta\phi}{\delta i}+\frac{\pi^{3}}{4\sqrt{2}}\left(\frac{\delta\phi}{\delta i}\right)^{3}+\frac{\pi^{5}}{4\sqrt{2}}\left(\frac{\delta\phi}{\delta i}\right)^{5}+\dots\label{eq:jav_expansion}
\end{equation}
as we anticipated from the qualitative reasoning of the previous paragraph
($\partial^{n}\jav/\partial\phi_{a}^{n}\propto(\delta i)^{-n}$).
We numerically find that this result remains qualitatively valid even when $\beta_{L}$
is not small.

\begin{figure}
\includegraphics[width=1\columnwidth]{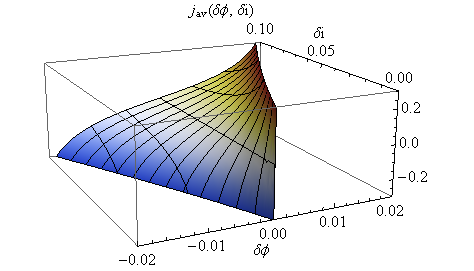}

\caption{\label{fig:The-average-circulating}(color online) The average circulating current
in the region $0<\delta i\ll 1$, as given by the analytical approximation
(see Eq.~\ref{eq:jav-approx}). The singular behavior at the cusp
point ($\delta i=0$, $\delta\phi=0$) is evident (\ref{eq:jav_expansion}).
This behavior can be exploited for the generation of highly nonlinear
response of the beam modes.}
\end{figure}

\subsection{Fundamental limits on the divergence of the Duffing coefficient}

The above discussion on the divergence of $\delta i$ disregards thermal
noise and $1/f$ noise. In reality, these noises render the limit
$\delta i\to0$ unphysical. First, we consider the limitation set
by thermal noise.  This can be accounted for by adding a white noise
term to the equation for $\gamma$. We then obtain a nonlinear Langevin
equation with a critical point of the marginal type~\cite{kubo_fluctuation_1973,colet_relaxation_1989,caceres_stochastic_1995}.
A simple dimensional analysis argument indicates that when $\delta i=0$,
a noise-induced transition from $\gamma_{c+}$ to $\gamma_{c-}$ should
occur on a time scale $\tau_{N}\propto(|v'''(\gamma_{c\pm})|^{2}|\Gamma)^{-\frac{1}{3}}$,
where $\Gamma=2\pi k_{B}T/I_{0}\Phi_{0}$ is the normalized diffusion
coefficient and $T$ is the junction temperature. For the above picture,
and in particular Eq.~(\ref{eq:jav_expansion}), to be correct, we therefore
require $\tau_{N}\gg\tau_{\pm}$, where $\tau_{\pm}$ is the time
spent near the critical points $\gamma_{c}$ (see appendix~\ref{sec:Analytical-expression-for-jav}).
This translates to a required operating temperature of $T\ll T_{{\rm max}}$,
where 
\begin{equation}
T_{{\rm max}}=\frac{2E_{J}}{k_{B}}\sqrt{\frac{\delta i^{3}}{\pi^{7}\beta_{L}}},\label{eq:max_temp}
\end{equation}
and $E_{J}$ is the junction energy. A more formal treatment that
leads to similar results, and shows that this is the relevant timescale
when $\delta i>0$ as well, can be found in~\cite{colet_relaxation_1989,caceres_stochastic_1995}.

Secondly, we consider the effect of $1/f$ fluctuations in the critical
current and flux. These two noise sources are an active area of current
research~\cite{van_harlingen_decoherence_2004,koch_model_2007,anton_low-frequency_2012,anton_magnetic_2013}
due to their crucial effect on superconducting qubit dephasing times.
Since our goal is to make a rough assessment of the limits of validity
of Eq.~(\ref{eq:jav_expansion}), we will consider only the order of
magnitude of these fluctuations. The most direct limitation on the
divergence in Eq.~(\ref{eq:jav_expansion}) is due to fluctuations in
$I_{0}$, which directly translate to fluctuations in $\delta i$.
Assuming that these fluctuations dominate those in the bias current,
and neglecting the noise input bandwidth due to its weak (logarithmic)
contribution to $\left\langle I_{0}^{2}\right\rangle $, we can use
the data in~\cite{anton_low-frequency_2012,van_harlingen_decoherence_2004}
to give the rough estimate $\sqrt{\left\langle \delta i{}^{2}\right\rangle }\simeq10^{-6}$.
The flux noise, following data reported in~\cite{koch_model_2007,anton_magnetic_2013},
can be estimated with roughly the same figure of $\sqrt{\left\langle \delta\phi^{2}\right\rangle }\simeq10^{-6}$.

Using Eq.~(\ref{eq:jav_expansion}), and the above considerations,
we see that the most stringent limitation comes from Eq.~(\ref{eq:max_temp}),
which implies that for a JJ with $I_{0}=100{\rm \mu A}$, $\beta_{L}=0.1$
and at $T=20{\rm mK}$, the deterministic dynamics outlined above
remain valid only when $\delta i\gtrsim0.015$. This sets an upper
bound on the size of the Duffing coefficient that can be obtained
in this device.

\section{Summary}

We have demonstrated that an interaction between a dc-SQUID and a
mechanical beam may be used to generate a nonlinearity in the beam
which is both strong and tunable. By decoupling the displacement detection
mechanism from the SQUID-beam system, we were able to characterize
the effective potential of the beam for the entire control parameter
space. The effective potential was calculated numerically, and a partial
agreement with experimental results was found. In a system with improved
operating parameters and beams that are close in frequency, many interesting
experiments such as two-mode noise squeezing~\cite{Xue_064305} and
thermally activated switching may be undertaken. Finally, it remains
an important question to consider whether operating a system close
to its bifurcation point may enable the experimenter to explore macroscopically
distinct quantum states that are inaccessible by other means.

\begin{acknowledgments}
The authors would like to thank J. M. Martinis for enlightening discussions.
This work was supported in part by the German Israel Foundation, in
part by the Israel Science Foundation, in part by the Bi-National
Science Foundation, in part by the Israel Ministry of Science, in
part by the Russell Berrie Nanotechnology Institute and in part by
the European STREP QNEMS Project.
\end{acknowledgments}

\bibliographystyle{apsrev}
\bibliography{squid_beam_nourl,squid_beam_new} %,\string"C:/Users/liorella/Dropbox/thesis/microwave resonator/optical_cavity_oscillator_nourl\string"}

\appendix

\section{\label{sec:Characterization-of-SQUID-beams}Characterization of SQUID
and beams}

\subsection{SQUID parameters}

\begin{figure}
\includegraphics[width=1\columnwidth]{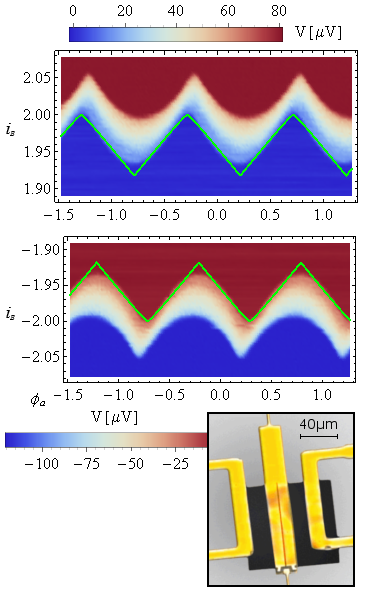}

\caption{(color online) SQUID voltage as a function of control parameters used to extract
its parameters in the absence of split-coil magnetic field\label{fig:Experimental-data_squid_params}.
The bias current axis is normalized such that $i_{B}=I_{B}/I_{0}$,
where $I_{0}=317.5{\rm \mu A}$. The extracted values, using the calculated
$i_{c\pm}(\phi_{a})$ curves (green lines), are $\beta_{L}=21.1$
and $\alpha_{I}=-0.027$, corresponding to $L=69{\rm pH}$, which
is consistent with the value calculated using 3D-MLSI~\cite{khapaev_3d-mlsi:_2001}.
Note the absence of a sharp cusp. The voltage is truncated at higher currents
due to voltage compliance settings. Inset: A false-color optical micrograph
of the device.}
\end{figure}

We fabricated a dc-SQUID with two nearly identical Nb/Al(AlOx)/Nb
Josephson junctions (JJs)~\cite{yuvaraj_fabrication_2011} in a washer
configuration (see inset in Fig.~\ref{fig:Experimental-data_squid_params}).
The SQUID was characterized in zero split-coil magnetic field. It
was found to have $I_{0}=(I_{0,1}+I_{0,2})/2=317.5{\rm \mu A}$ at
zero magnetic field and at temperature $T=3.81{\rm K}$, and $I_{0}=30\pm2{\rm \mu A}$
with $B=60{\rm mT}$. The self inductance parameter is $\beta_{L}=2L_{0}I_{0}/\Phi_{0}=21.1$
at zero field, where $L_{0}=69{\rm pH}$ is the loop inductance when
the beams are in their equilibrium positions. From the frequency response
of the beams, we observed that this parameter was reduced to $\beta_{L}=4$
in the magnetic field. The critical current asymmetry was found to
be $\alpha_{I}=(I_{0,2}-I_{0,1})/2I_{0}=-0.027$. Since the voltage
response of the SQUID was non-hysteretic, we determined that $\beta_{c}=2\pi I_{0}R^{2}C/\Phi_{0}<1$
at zero field, where $C$ is the equivalent junction capacitance and
$R\simeq1\Omega$ is the equivalent junction shunt resistance. In
practice, $\beta_{c}$ could be neglected in our analysis. The noise
coefficient is $\Gamma=k_{B}T/E_{J}=5\times10^{-3}$ when the magnetic
field is turned on. Here $T$ is the junction temperature and $E_{J}=I_{0}\Phi_{0}/2\pi$
is the junction energy. The oscillation frequency of the JJs is $\omega_{c}=2\pi \times 14.7{\rm GHz}$
with applied magnetic field.

The inductance $L_{0}$ of the SQUID loop was calculated using a numerical
software (3D-MLSI~\cite{khapaev_3d-mlsi:_2001}). The parameters
$\beta_{L}$ and $\alpha_{I}$ were extracted by measuring the voltage
as a function of control parameters, which provided the $i_{c\pm}(\phi_{a})$
curves that separate the S-state from the R-state for positive and
negative bias currents, respectively. (See Fig.~\ref{fig:Experimental-data_squid_params}).
Note that in contrast to the theoretical prediction and early SQUID
measurements~\cite{de_waele_critical_1969}, our SQUID did not show
a sharp cusp point at the points of minimal $|i_{c\pm}|$.

The mutual inductance between the SQUID and the flux line is $M=1.88{\rm pH}$.
The strength $60{\rm mT}$ of the applied split-coil magnetic field
was calculated both analytically and using finite elements analysis,
with results agreeing within $95\%$. We finally remark that no shunting
resistance was required in order to overdamp the SQUID. This is possibly
due to conducting channels created at the junction barrier during
the junction sculpting process with the focused ion beam (FIB)~\cite{yuvaraj_fabrication_2011}.

\subsection{Mechanical parameters}

Each of the doubly-clamped beams has length $\ell=100{\rm \mu m}$,
lateral width $w=14{\rm \mu m}$, thickness $t={\rm 0.7\mu m}$ and
bare mass $m=8.7{\rm ng}$, with $m_{{\rm eff}}=0.735m$~\cite{ekinci_ultimate_2004}.
The mode frequencies of the beams were characterized at zero magnetic
field, and only the lowest frequency mode was actuated. The mode profile
(measured by scanning the position of the optical fiber) indicated
that only one of the beams vibrated with this frequency, and that
the second beam had a much higher fundamental flexural mode of $f_{1}=673.5{\rm KHz}$,
so that intermode coupling could be safely disregarded.

\subsection{\label{sub:Coupling-constant-g}Coupling constant}

In this section we discuss the coupling constant $g=\lambda\ell B\sqrt{I_{0}/m_{{\rm eff}}\Phi_{0}}$ between the SQUID and the beam. Here, $\lambda$ is a
geometric correction factor which includes corrections due to mechanical
mode shape, effective mass mode and magnetic field screening. To extract
$\lambda$ from the measurements, we use the fact that for $i_{B}<\icmin$,
the Lorentz force acting on the beam is nearly linear in $u$ for
almost all values of $\phi_{a}$(see Fig.~\ref{fig:Frequency-response-s-state}).
This translates to a nearly constant shift of the frequency of the
mechanical mode, which we can use to fit $\lambda$. From this we
obtain $\lambda=0.6$.

\subsection{Detection and actuation}

Capacitive actuation and detection of the mechanical mode are both
accomplished using the Niobium coated optical fiber, which is connected
galvanically to the output of a sweeping function generator. The function
generator also provides a reference signal to an RF lock in amplifier
(LIA). Since the SQUID is top-coated with gold, it is highly reflective,
and forms one side of an optical cavity. The other side of the cavity
is formed at the dielectric interface between the tip of the fiber
and free space. The power reflected from this optical cavity is converted
to voltage by an RF photodetector, and fed to the input of the LIA.
In this manner, the LIA functions as a network analyzer with the capability
to sweep the driving frequency both in the up and down directions.
This two-sided sweep is required in order to characterize the bistable
regions in the frequency response of the beam.

\section{\label{sub:Modeling-the-SQUID-beam-1}Modeling the SQUID-beam interaction}

The normalized equations of motion for a symmetric SQUID in the RCSJ
model and the amplitude of the driven mode in the harmonic approximation
are\begin{widetext}\begin{subequations}\label{eq:EOMS-normalized}
\begin{eqnarray}
\beta_{c}\omega_{c}^{-2}\ddot{\gamma}+\omega_{c}^{-1}\dot{\gamma}+2\cos\left(\frac{\gamma_{-}}{2}\right)\sin\left(\frac{\gamma}{2}\right) & = & i_{B}+i_{N,+}\label{eq:norm_gp}\\
\beta_{c}\omega_{c}^{-2}\ddot{\gamma}_{-}+\omega_{c}^{-1}\dot{\gamma}_{-}+2\cos\left(\frac{\gamma}{2}\right)\sin\left(\frac{\gamma_{-}}{2}\right) & = & -2j+i_{N,-}\label{eq:norm_gm}\\
\frac{\gamma_{-}}{2\pi}-\phi_{a}-u & = & \frac{1}{2}\beta_{L}j\label{eq:norm_j}\\
\ddot{u}+Q_{m}^{-1}\omega_{0}\dot{u}+\omega_{0}^{2}u & = & g^{2}\left(\frac{1}{2}i_{B}+j\right)+h_{d}\cos(\omega_{p}t),\label{eq:norm_ym}
\end{eqnarray}
\end{subequations}\end{widetext}where $\gamma_{-}=\gamma_{2}-\gamma_{1}$,
$i_{N,\pm}=I_{N,\pm}/I_{0}$, and $I_{N,\pm}$ is current noise in
the junctions. The response of the driven mode to the excitation by
the SQUID was obtained by calculating $\jav$, as defined in the body
of the text, for the range $0<i_{B}<3$, $0<\phi_{a}<1$ of the control
parameters. When the SQUID was in the S-state, $\jav$ was obtained
by finding all roots of Eq.~(\ref{eq:norm_gp}-\ref{eq:norm_j}) in the
steady state. In general, more than one such root (or well of the
SQUID potential) exists when $\beta_{L}>0$. However, this multiplicity comes into effect only near values of $\phi_{a}$ for which a well disappears (see theoretical panel
in Fig.~\ref{fig:Frequency-response-s-state}), which are the points
near which discrepancy between the model and the experiment exists.
In the R-state, $\jav$ was found by integrating $j(t)$ which was
numerically computed using Eq.~(\ref{eq:norm_gp}-\ref{eq:norm_j}) over
a single period $\Theta$. The asymmetry was found to be small in
our device ($\alpha_{I}=-0.027$), and therefore was not taken into
account in the numerical calculations.

After $\jav(\phi_{a},i_{B})$ was obtained, the derivatives $\partial\jav/\partial\phi_{a}$,
$\partial^{3}\jav/\partial\phi_{a}^{3}$ were calculated numerically.
These were used to obtain the frequency shift and Duffing coefficient
for the equation of the mode amplitude in the rotating wave approximation~\cite{landau_mechanics_1976,lifshitz_nonlinear_2008}
\begin{equation}
\left[\left(\delta+\frac{1}{2}\epsilon d_{1}+\frac{3}{8}\epsilon d_{3}|A|^{2}\right)^{2}+\left(\frac{1}{2Q_{m}}\right)^{2}\right]|A|^{2}=\frac{1}{4}\epsilon_{d}^{2},
\end{equation}
where $\delta=(\omega_{p}-\omega_{0})/\omega_{0}$, $d_{1}=\frac{1}{2}\partial\jav/\partial\phi_{a}$,
$d_{3}=\frac{1}{6}\partial^{3}\jav/\partial\phi_{a}^{3}$, $\epsilon_{d}=h_{d}/\omega_{0}^{2}$,
$\epsilon=g^{2}/\omega_{0}^{2}$, and $u(t)=\frac{1}{2}Ae^{-i(1+\delta)\omega_{0}t}+{\rm c.c}$.
This was used to generate the theoretical panels in Figs.~\ref{fig:Comparison-of-frequecy-low}, \ref{fig:Comparison-of-frequecy-hi} and~\ref{fig:Frequency-response-s-state}.

\section{\label{sec:Analytical-expression-for-jav}Analytical expression for
$\jav$ near the bifurcation cusp point}

\begin{figure}
\includegraphics[width=1\columnwidth]{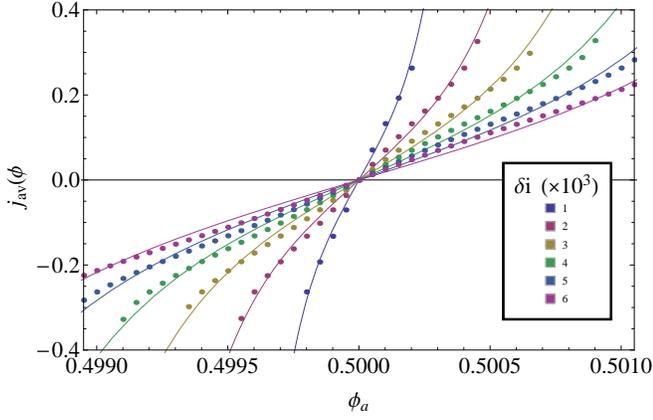}

\caption{\label{fig:A-comparison-of-analytic-numeric}(color online) A comparison of the analytical
expansion Eq.~(\ref{eq:jav-approx}) and the value of $j_{{\rm av}}(\delta\phi)$
obtained by numerically integrating the equations of motion~(\ref{eq:EOMS-normalized}),
with $\beta_{L}=0.1$, $\beta_{c}=0.05$, no noise and $u=0$.}
\end{figure}

Following the main text, we expand the potential Eq.~(\ref{eq:gamma_pot})
around the points $\gamma_{c\pm}$ defined by $v''(\gamma_{c\pm})=0$
and $v'''(\gamma_{c\pm})<0$. Two such points exist for a single period
of $v(\gamma)+i_{B}\gamma$, and we find that near them the equation
of motion for $\gamma$ can be written as
\begin{equation}
\frac{{\rm d}\gamma}{{\rm d}\tau}=\delta i+c_{0\pm}+c_{2\pm}(\gamma-\gamma_{c\pm})^{2}+\dots,\label{eq:eoms_expanded-1}
\end{equation}
where\begin{subequations}
\begin{align}
c_{0\pm}(\delta\phi) & =\frac{\pi}{2}\beta_{L}\mp2\pi\left(\delta\phi+\frac{1}{2}\beta_{L}j_{c\pm}\right)\sqrt{1-j_{c\pm}^{2}},\\
c_{2\pm}(\delta\phi) & =\pm\frac{\pi}{4}\left(\delta\phi+2\beta_{L}j_{c\pm}\right)\sqrt{1-j_{c\pm}^{2}},
\end{align}
\end{subequations}and
\begin{equation}
j_{c\pm}(\delta\phi)=\pm\sqrt{\frac{1}{2}+\left(\frac{\delta\phi}{2\beta_{L}}\right)^{2}}-\frac{\delta\phi}{2\beta_{L}}.
\end{equation}
The solution of Eq.~(\ref{eq:eoms_expanded-1}) truncated after the quadratic
term is $\gamma_{\pm}(\tau)=\gamma_{c\pm}+\eta_{\pm}\tan\left(\pi\frac{\tau}{\tau_{\pm}}\right)$,
where $\eta_{\pm}$ and $\tau_{\pm}$ are given by

\begin{equation}
\tau_{\pm}=\frac{\pi}{\sqrt{\left(\delta i+c_{0\pm}\right)c_{2\pm}}}\label{eq:tau_pm-1}
\end{equation}
and
\begin{equation}
\eta_{\pm}=\sqrt{\frac{\delta i+c_{0\pm}}{c_{2\pm}}}.
\end{equation}

Since for $\delta\phi=0$ we have $j_{c\pm}=\pm1/\sqrt{2}$ and therefore
$c_{0\pm}=0$, we see that the time spent near the slow points indeed
scales as $(\delta i)^{-\frac{1}{2}}$, as expected from an infinite
period bifurcation~\cite{strogatz_nonlinear_2001}. We can now calculate
$\jav$ using these solutions and the fact that $j(\tau)=-\cos\left(\gamma/2\right)$,
and we obtain

\begin{align}
\jav & =\frac{1}{\Theta}\int_{0}^{\Theta}j(\tau){\rm d}\tau=\frac{j_{c+}\tau_{+}e^{-\frac{1}{2}\eta_{+}}+j_{c-}\tau_{-}e^{-\frac{1}{2}\eta_{-}}}{\tau_{+}+\tau_{-}}\nonumber \\
 & +O(\sqrt{\delta i}).\label{eq:jav-approx}
\end{align}
A comparison between $\jav$ obtained with this approximation and
the one calculated numerically using the equations of motion Eq.~(\ref{eq:norm_gp}-\ref{eq:norm_j})
is found in Fig.~\ref{fig:A-comparison-of-analytic-numeric}.

\end{document}